\documentclass[a4paper,fleqn]{cas-sc}
\pdfoutput=1 

\usepackage[authoryear,longnamesfirst]{natbib}

\def\tsc#1{\csdef{#1}{\textsc{\lowercase{#1}}\xspace}}
\tsc{WGM}
\tsc{QE}
\tsc{EP}
\tsc{PMS}
\tsc{BEC}
\tsc{DE}
\definecolor{colrev}{rgb}{0, 0, 0} 

\begin{document}
\let\WriteBookmarks\relax
\def\floatpagepagefraction{1}
\def\textpagefraction{.001}
\shorttitle{Moral Decisions in the Age of COVID-19: Your Choices Really Matter}
\shortauthors{F. Donnarumma and G. Pezzulo}

\title [mode = title]{Moral Decisions in the Age of COVID-19: \\Your Choices Really Matter}




\author[1]{Francesco Donnarumma}[orcid=0000-0003-4248-5360]
\cormark[1]
\ead{francesco.donnarumma@istc.cnr.it}


\address[1]{Institute of Cognitive Sciences and Technologies, National Research Council, Via S. Martino della Battaglia 44, 00185 Rome, Italy}

\author[1]{Giovanni Pezzulo}[orcid=0000-0001-6813-8282]
\ead{giovanni.pezzulo@istc.cnr.it}

\cortext[cor1]{Corresponding author}

\nonumnote{The authors declare that they have no conflict of interest.}

\begin{abstract}
The moral decisions we make during this period, such as deciding whether to comply with quarantine rules, have unprecedented societal effects. We simulate the ``escape from Milan'' that occurred on March 7th-8th \textcolor{colrev}{2020}, when many travelers moved from a high-risk zone (Milan) to southern regions of Italy (Campania and Lazio) immediately after an imminent lockdown was announced. Our simulations show that fewer than 50 active cases might have caused the sudden spread of the virus observed afterwards in these regions. The surprising influence of the actions of few individuals on societal dynamics challenges our cognitive expectations -- as in normal conditions, collective dynamics are rather robust to the decisions of few ``cheaters''. This situation therefore requires novel educational strategies that increase our awareness and understanding of the unprecedented effects of our individual moral decisions.
\end{abstract}




\maketitle

\section{Introduction}

During this COVID-19 pandemic, we are all required to make important moral decisions \citep{greene2013moral}. Our leaders have to make complex choices when setting up lockdown measures, which involve a trade off between potential benefits (e.g., saving more lives and avoiding a collapse of health care services) and costs (e.g., economic costs).
Yet not just our leaders, but also we as individuals make important moral decisions, such as whether or not to stick to lockdown or quarantine rules \citep{harris1995there,childress2002public,alkire2004global}. Indeed, lockdowns and quarantines have been described as altruistic acts to protect others, and especially elderly and fragile persons.
In standard conditions, a minority of “cheaters” who break the rules (e.g., when voting, paying taxes or breaking traffic rules) can be tolerated and such behavior does not significantly affect collective dynamics \citep{brennan2012ethics}. Conversely, in this pandemic situation, even the choice of a few individuals may matter at the collective level -- for better or for worse.

\section{The effects of the ``escape from Milan''}

To illustrate the unprecedented effect of the choices of few individuals at the collective level, we model a paradigmatic case of moral decision: the ``escape from Milan'' that occurred in the early days of the COVID-19 spread in Italy. On March 7th, there were rumors of an imminent lockdown in Lombardia, which had a large number of COVID-19 cases. \textcolor{colrev}{During the weekend of March 7th-8th, several media sources documented a massive ``escape'' from Milan (the biggest city of Lombardia): the fleeing of people to southern regions of Italy, such as Campania and Lazio, where few cases of COVID-19 had been reported. This situation was extensively covered and stigmatized in the Italian public debate, with public authorities and media accused of disclosing critical information too early and travelers accused of spreading the virus in southern regions of Italy, whose health care services were unprepared. It was subsequently argued later that the public perception of an actual ``escape'' (or exodus, at it was also called) was exaggerated, but still the number of people leaving Lombardia for southern regions was in the order of a few thousands -- whereas it declined drastically immediately afterwards, as an effect of the lockdown \citep{BERIA2021102616}.}

\begin{figure}
\centering
\LARGE
\textbf{(A)}    \hspace{8cm}               \textbf{(B)}\\
\hspace{-1.5em}\includegraphics[width=.52\linewidth]{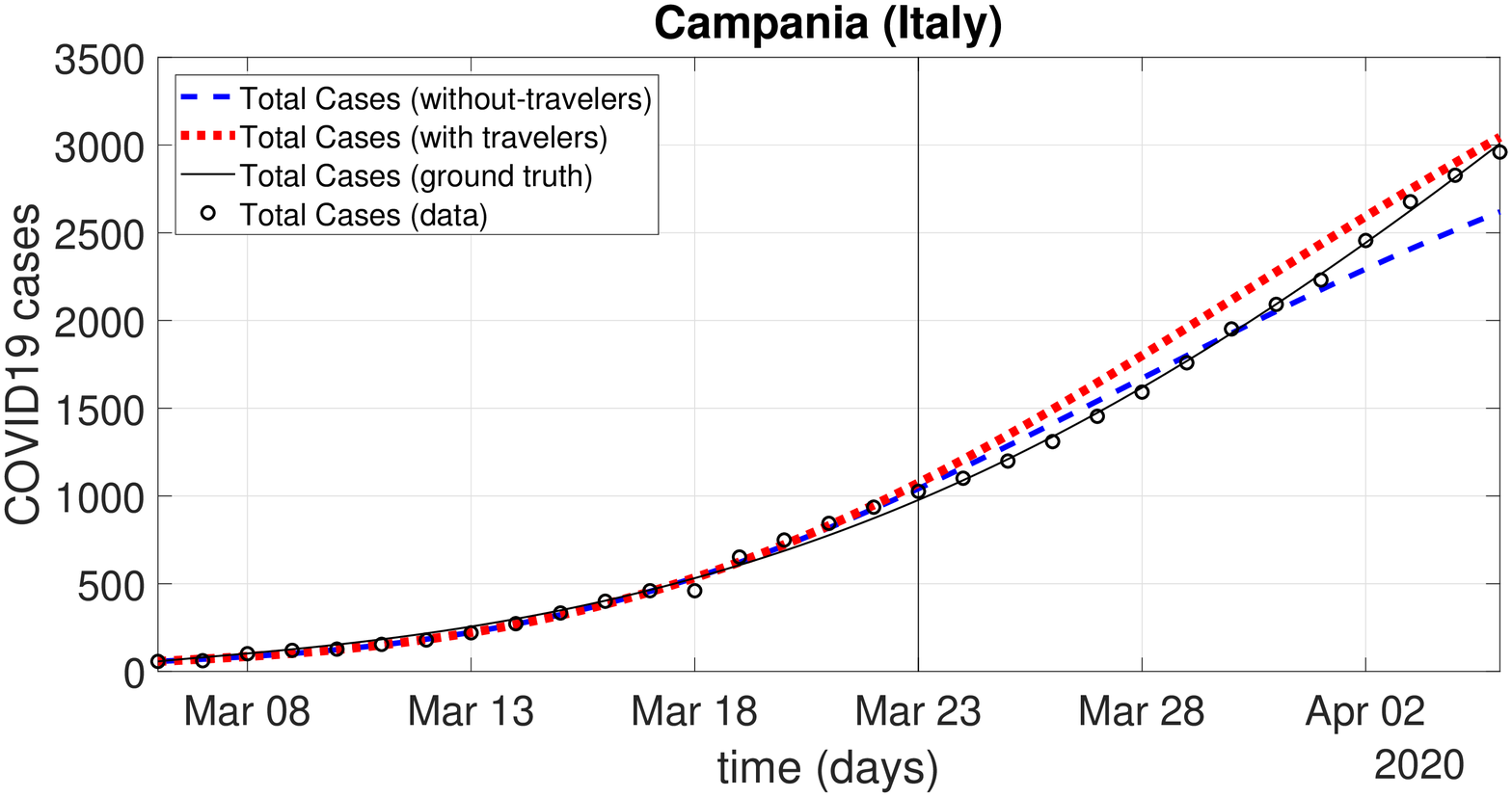} \includegraphics[width=.52\linewidth]{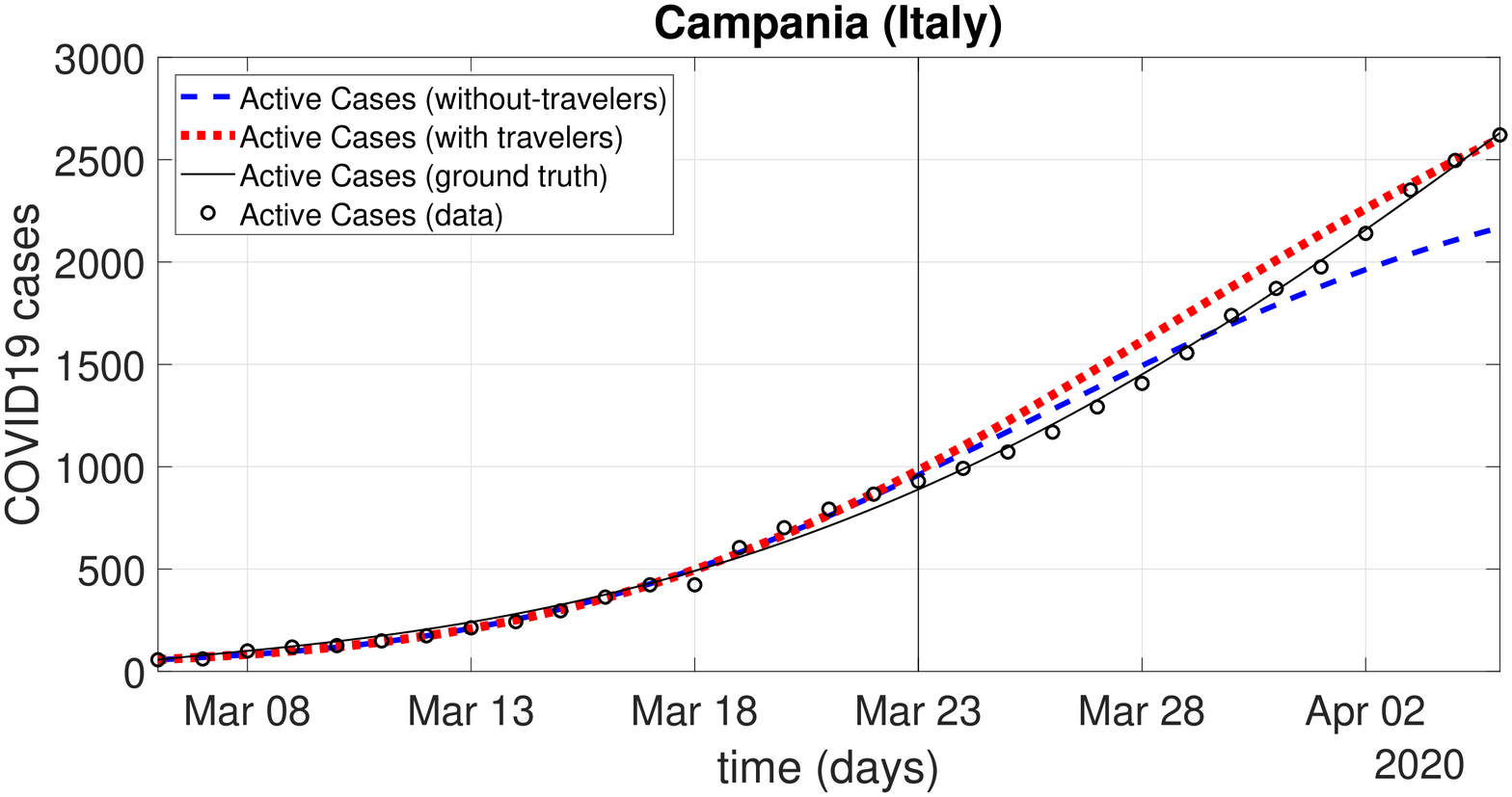}\\

\textbf{(C)} \hspace{8cm}            \textbf{(D)}\\
\hspace{-1.5em}\includegraphics[width=.52\linewidth]{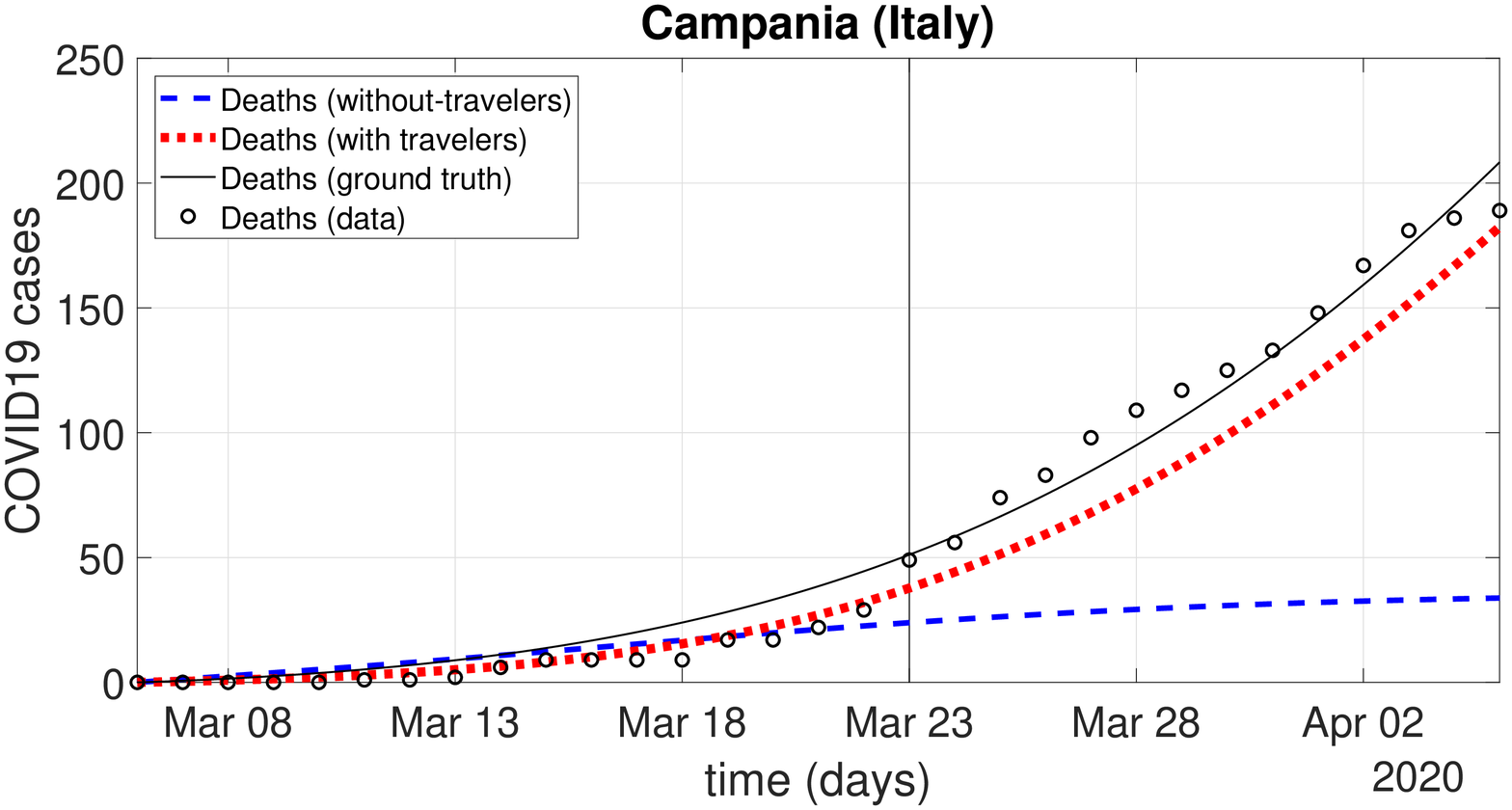}
\includegraphics[width=.52\linewidth]{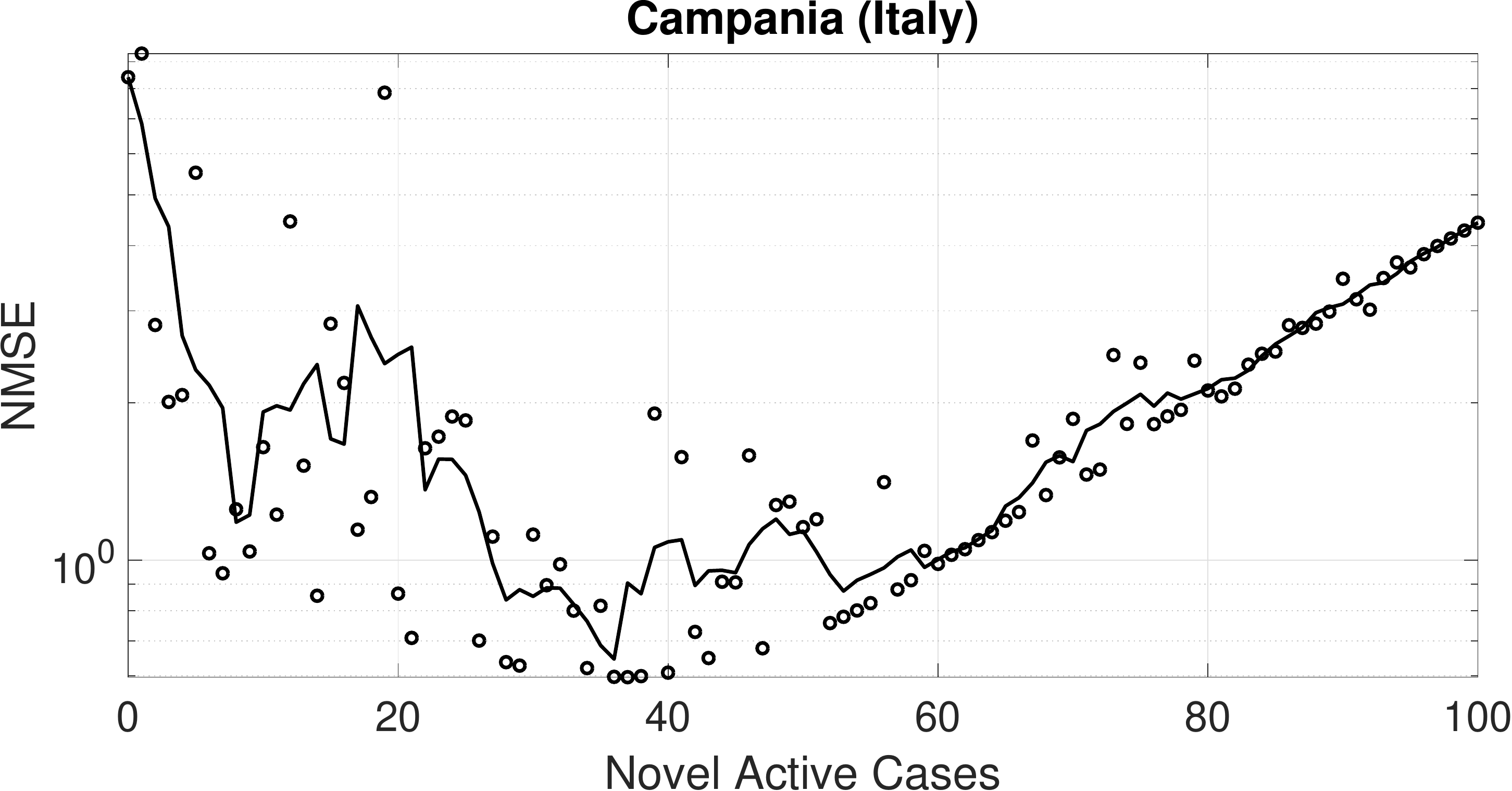}\\  

\LARGE
\caption{\label{fig:CampaniaSim}
Simulations of COVID-19 spread in Campania (A-C). Black circles are real data and lines are fits from different instances of the SEIR model. Plots show Simulations of total cases (Panel (A)), active cases (Panel (B)) and deaths (Panel (C)), using  \emph{without-travelers} (blue dashed lines),  \emph{with-travelers} (red dotted lines) and \emph{ground-truth} models (black solid lines). \emph{Without-travelers} and \emph{with-travelers} models are trained using data before the vertical line and used to predict the next days. \emph{Ground-truth} model is trained with all available data. Panel (D) Reconstruction error (NMSE) of \emph{with-travelers} model with a varying number of novel active cases (1 to 100). The black circles represent the NMSE values obtained by adding 1 to 100 novel active cases; whereas the black continuous line is a 5-points moving average performed on NMSE values. The NMSE values are computed by comparing the output of the simulation (from March 23th to April 5th) and the corresponding real data of deaths, active and total cases.} 

\end{figure}

\begin{figure}
\centering
\LARGE
\textbf{(A)}    \hspace{8cm}               \textbf{(B)}\\
\hspace{-1.5em}\includegraphics[width=.52\linewidth]{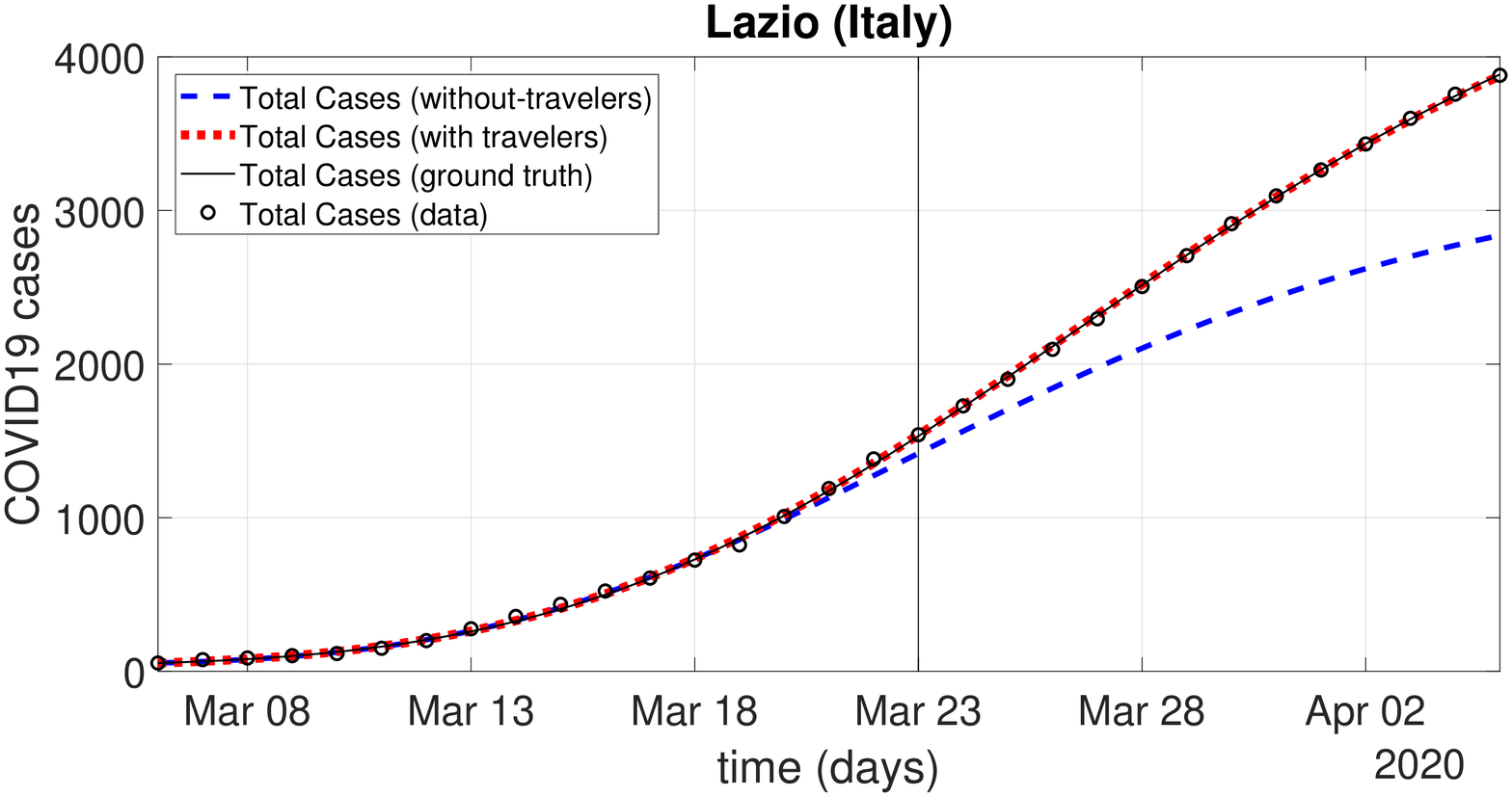} \includegraphics[width=.52\linewidth]{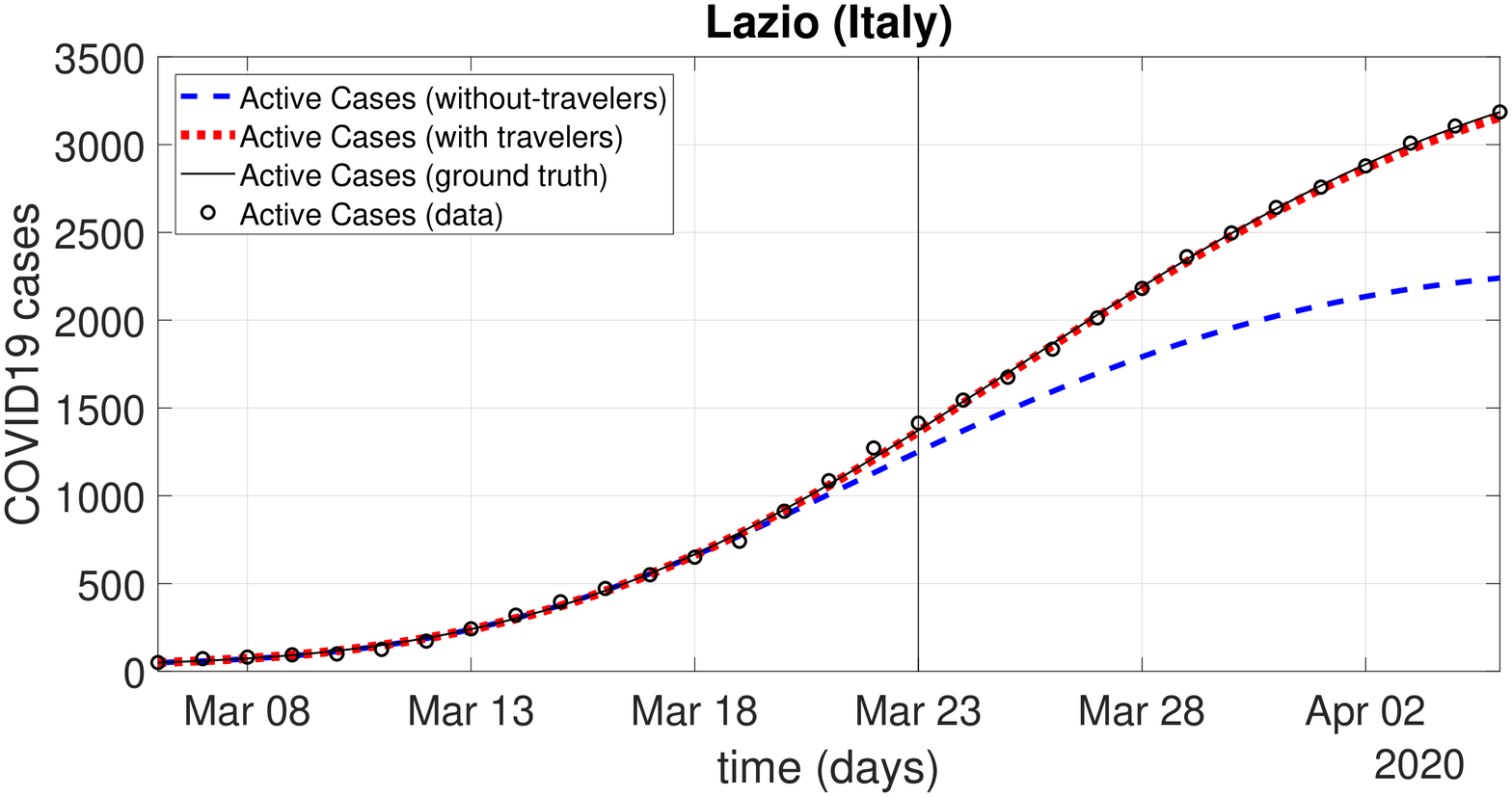}\\

\textbf{(C)} \hspace{8cm}            \textbf{(D)}\\
\hspace{-1.5em}\includegraphics[width=.52\linewidth]{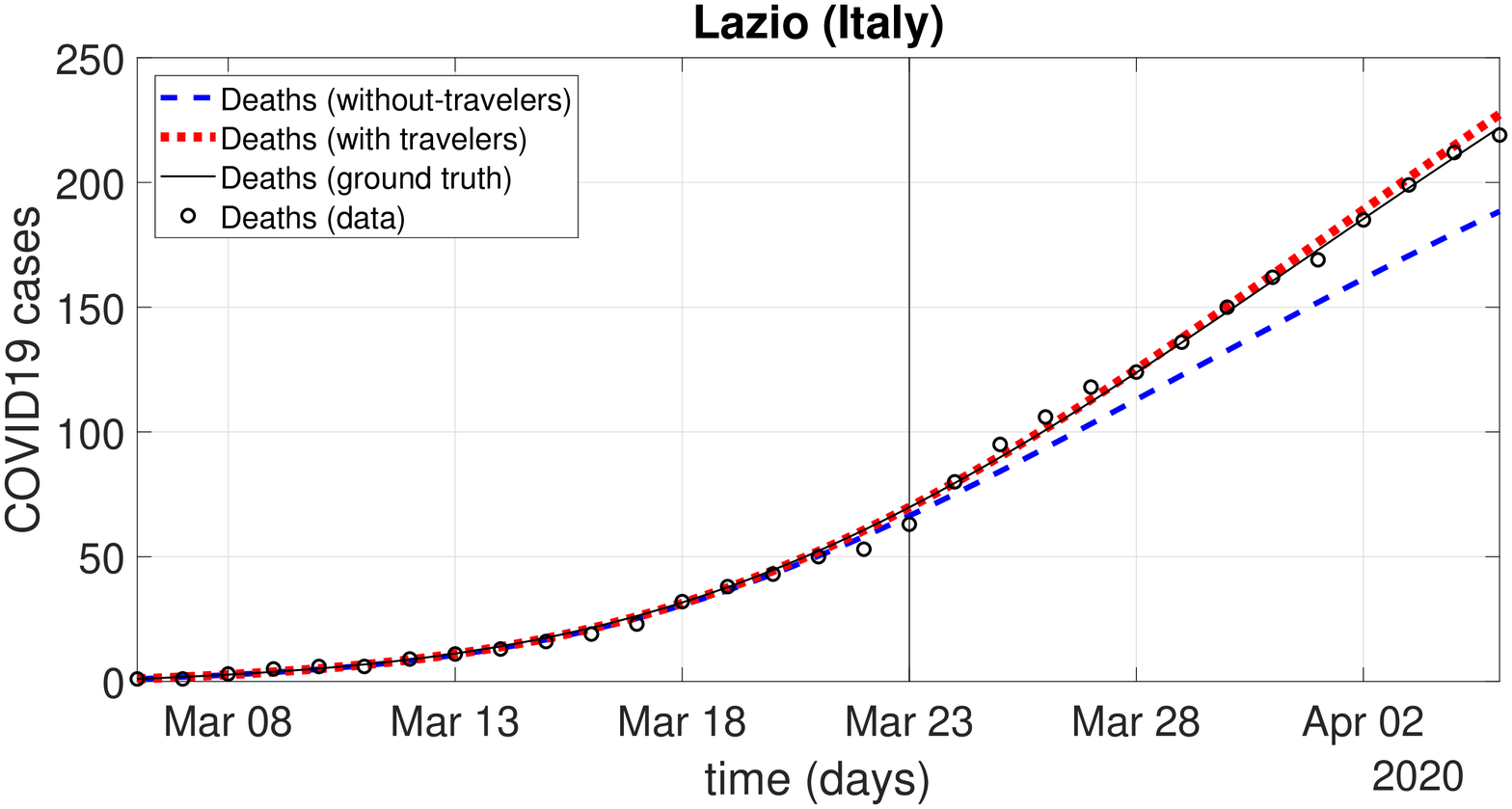}
\includegraphics[width=.52\linewidth]{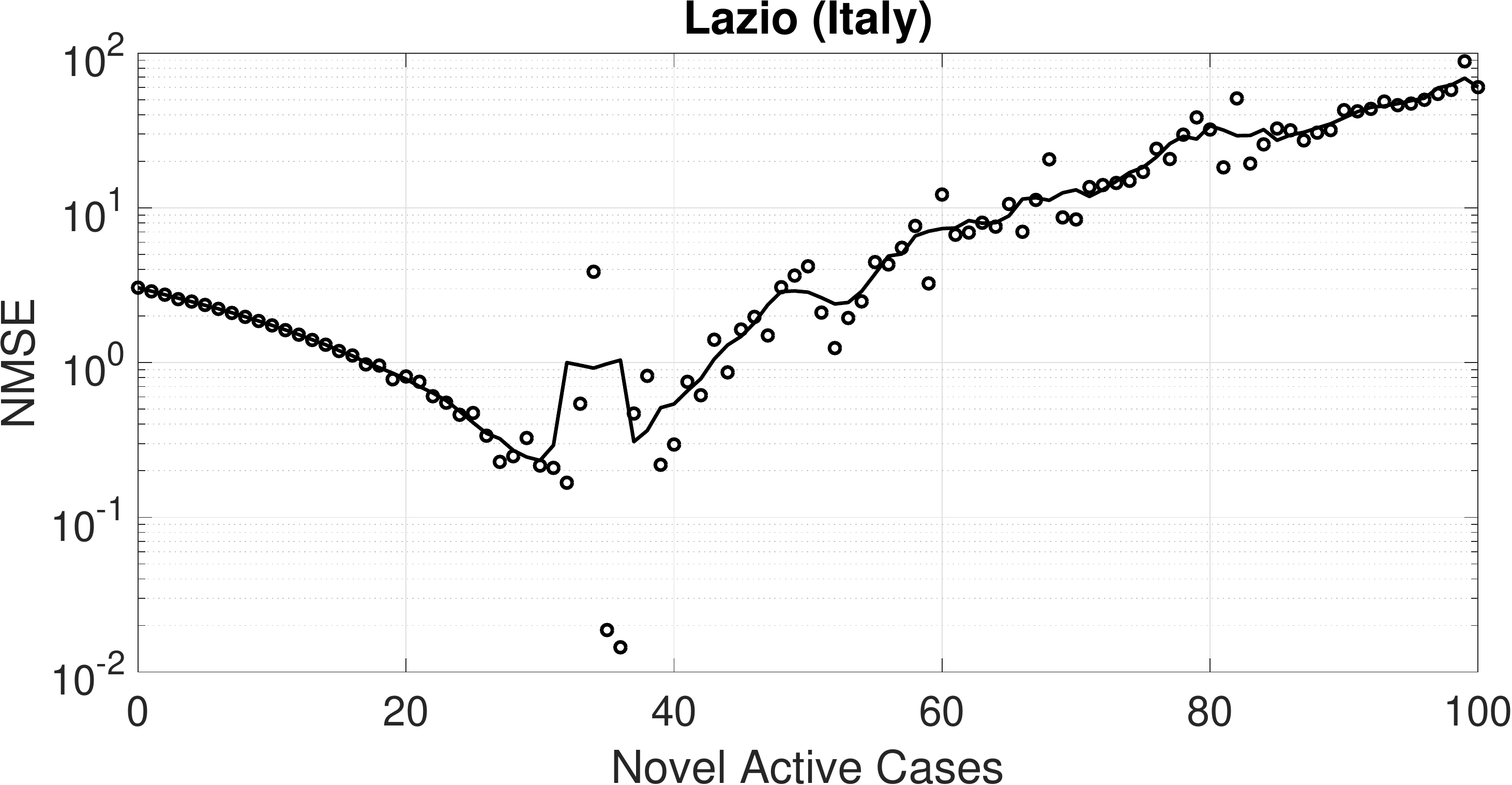}\\  

\LARGE
\caption{\label{fig:LazioSim}
Simulations of COVID-19 spread in Lazio (A-C). Black circles are real data and lines are fits from different instances of the SEIR model. Plots show simulations of total cases (Panel (A), active cases (Panel (B)) and Deaths (Panel (C)), using  \emph{without-travelers} (blue dashed lines),  \emph{with-travelers} (red dotted lines) and \emph{ground-truth} models (black solid lines). \emph{Without-travelers} and \emph{with-travelers} models are trained using data before the vertical line and used to predict the next days. \emph{Ground-truth} model is trained with all available data. Panel (D) Reconstruction error (NMSE) of \emph{with-travelers} model with a varying number of novel active cases (1 to 100). The black circles represent the NMSE values obtained by adding 1 to 100 novel active cases; whereas the black continuous line is a 5-points moving average performed on NMSE values. The NMSE values are computed by comparing the output of the simulation (from March 23th to April 5th) and the corresponding real data of deaths, active and total cases.} 

\end{figure}

\begin{figure}
\centering
\LARGE
\textbf{(A)}    \hspace{8cm}               \textbf{(B)}\\
\hspace{-1.5em}\includegraphics[width=.52\linewidth]{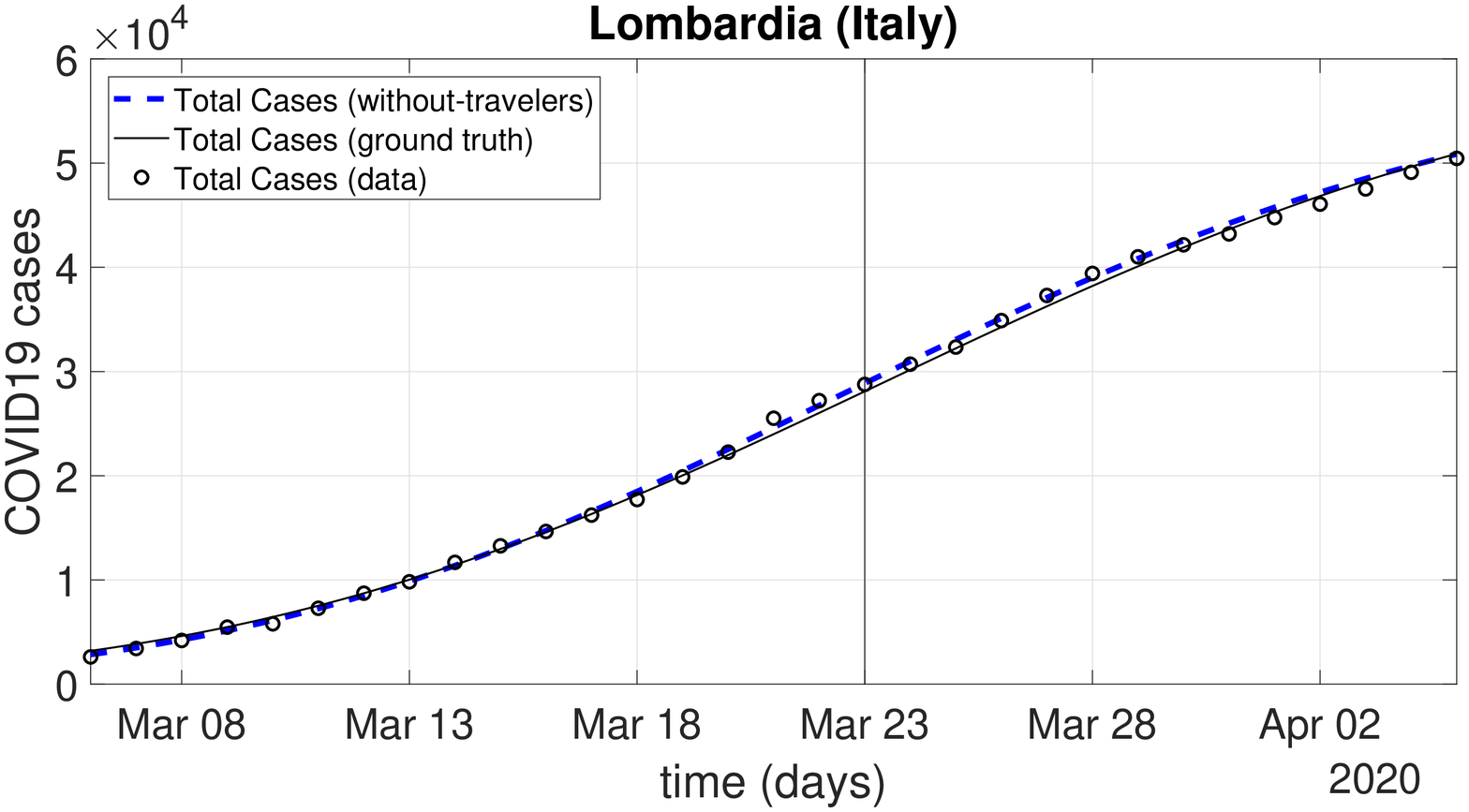} \includegraphics[width=.52\linewidth]{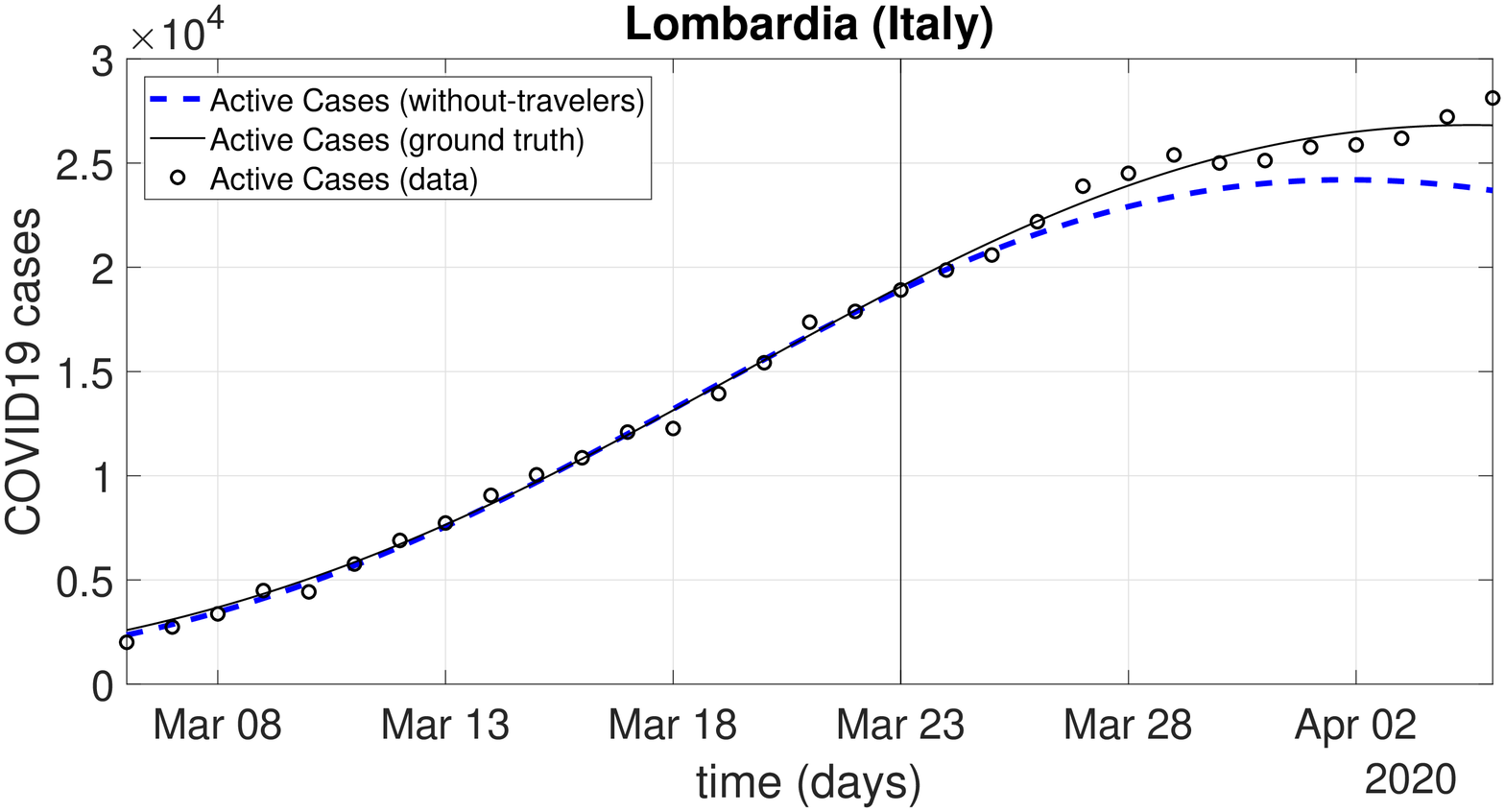}\\

\textbf{(C)}
\\
\hspace{-1.5em}
\includegraphics[width=.52\linewidth]{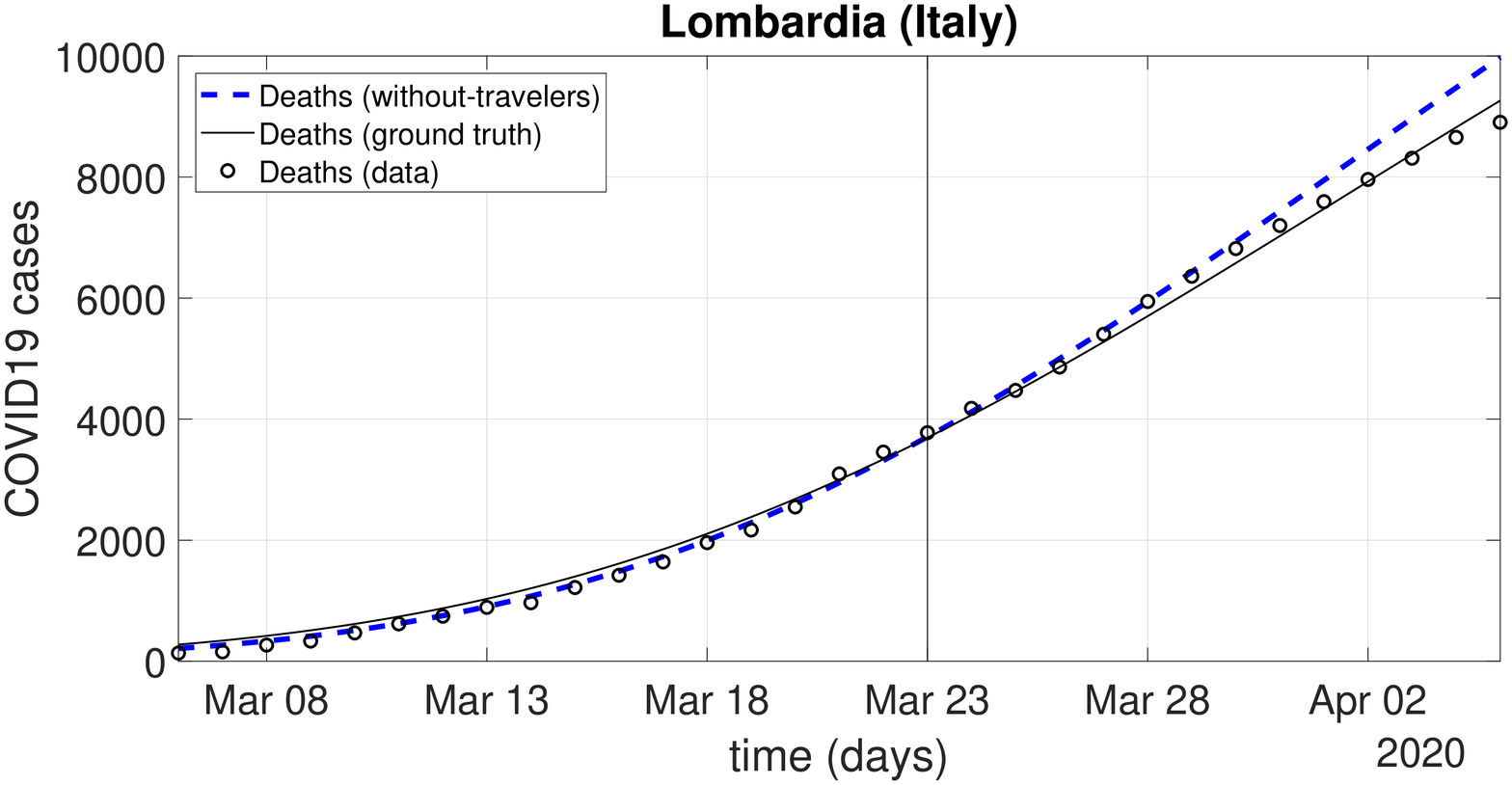}\\  

\LARGE
\caption{\label{fig:LombardiaSim}
Simulations of COVID-19 spread in Lombardia (A-C). Black circles are real data and lines are fits from different instances of the SEIR model. Plots show simulations of total cases (Panel (A), active cases (Panel (B)) and Deaths (Panel (C)), using  \emph{without-travelers} (blue dashed lines) and \emph{ground-truth} models (black solid lines). \emph{Without-travelers} and model is trained using data before the vertical line and used to predict the next days. \emph{Ground-truth} model is trained with all available data.} 

\end{figure}

COVID-19 cases in Campania and Lazio started to increase more steeply during the period of time between March 18th and March 23rd -- about two weeks after the escape from Milan, and despite the lockdown in these regions. We tested \textcolor{colrev}{(see Figures \ref{fig:CampaniaSim}, \ref{fig:LazioSim}, \ref{fig:LombardiaSim} and Materials and Methods)} whether this trend was coherent with the previous pattern in the same regions (hypothesis 1) or incoherent, hence suggesting possible external factors such as the arrival of ``travelers'' from other regions who were active COVID-19 cases (hypothesis 2). 
To adjudicate between the two hypotheses, we trained a model with data from Feb 24th to March 23rd, simulating the next days in the same region. \textcolor{colrev}{This model (\emph{without-travelers}, see Figures \ref{fig:CampaniaSim}A-B-C, \ref{fig:LazioSim}A-B-C, and \ref{fig:LombardiaSim}A-B-C  blue dashed lines) severely underestimates total cases, active cases and deaths from March 24th to April, 5th suggesting a change in trend.}

To assess whether this change in trend could have been caused by a small number of ``travelers'' from other regions, we performed another simulation \textcolor{colrev}{(\emph{with-travelers}, see Figures \ref{fig:CampaniaSim}A-B-C and \ref{fig:LazioSim}A-B-C, red dotted lines)} using the same data as before and adding a varying number of novel active cases to the model.
Our results show that a very small number of novel cases can be used to accurately predict the number of total cases, active cases and deaths from March 24th to April 7th (best fit with 37 novel cases for Campania and 36 for Lazio, \textcolor{colrev}{see Figures \ref{fig:CampaniaSim}C and \ref{fig:LazioSim}C).} Thus, very few novel cases are sufficient to explain the changed COVID-19 trend occurred around March 18th - 23th, i.e., about two weeks after the massive escape from Milan. This novel trend had dramatic effects: e.g. in Campania up until April 5th, there were 341 additional total cases and 155 additional deaths compared to what could have happened without $37$ novel cases (compare solid and dashed lines).
Conversely, a simulation of Lombardia does not show any change of trend \textcolor{colrev}{(Figure \ref{fig:LombardiaSim}).} 

In sum, our results show that a few novel cases from outside regions (potentially including travelers from Milan) may have caused a change in the trend of virus expansion in Campania and Lazio -- lately mitigated by lockdown measures. \textcolor{colrev}{Note that our analysis suggests that the “escape from Milan” is \emph{sufficient} to explain the exponential spread of the virus in southern regions, not (or not necessarily) that it actually caused it. Indeed, there are other scenarios that could potentially explain the same phenomenon. For example, it is possible that there were already independent clusters in other regions -- coherent with evidence of multiple independent clusters of the virus in Italy \citep{di2020genomic}. Furthermore, it is possible that the spread of the virus was due to travelers from other regions of the world, since the restrictions to enter Italy at that time were not severe. However, the goal of our simulations was not to prove a causal relation between the “escape from Milan” and the spread of the virus, but rather to highlight that very few cases from outside regions may have been enough to increase the virus exposure to disastrous levels in the general population -- and hence to draw attention to the fact that people's individual moral choices really do matter in the context of societal crises.}

\textcolor{colrev}{In other words, we used this example to illustrate how in some conditions, such as the pandemic, the choices of a small number of individuals can have significant effects at the collective level.} This situation is very uncommon in the moral decisions we usually make -- and demands increased awareness. Are we aware that our moral decisions these days can be so impactful?

\section{The COVID-19 pandemic exposes the limits of our adaptive rationality}

\textcolor{colrev}{When deciding whether or not to leave Milan, potential travelers have to balance their own benefits with potential dangers for the safety of themselves, their families and other residents of southern regions. Our intention here is not to stigmatize the choices of those who decided to leave Milan or to argue that they acted selfishly: they could have had opportunistic reasons as well as reasons that we perceive to be morally legitimate; for example, to care for an elderly relative \citep{van2019computational,schiffer2021moral}. However, even people who left for morally laudable reasons may have not correctly assessed the risks and benefits of their decision and hence failed to realize the potential societal costs of their choices. This is a cognitive problem that needs to be addressed through future interventions.}

\color{black}

Indeed, the COVID-19 pandemic poses significant cognitive challenges to our ability to make adaptive decisions. The impending sense of danger, stress, urgency and isolation that we all face, together with an over-exposure to multiple and sometimes conflicting sources of information, all contribute to creating a challenging context to deploy our cognitive skills. \textcolor{colrev}{Decision-making in such critical conditions is notoriously difficult, given the lack of prior experience and the high risks at stake \citep{shortland2020choice}}. 

These problems are exacerbated by some of our so-called cognitive limitations and biases. We are unable to correctly understand and predict phenomena that grow exponentially (\emph{exponential growth bias} \citep{levy2017exponential,wagenaar1975misperception}). 
We also tend to steeply discount (positive or negative) events that occur after some temporal delay, or at some spatial distance from us \citep{berns2007intertemporal,calluso2015analysis}; and we tend to disregard information that is not compatible with our prior beliefs (\emph{confirmation bias} \citep{plous1993psychology}) \textcolor{colrev}{-- perhaps because ``changing mind'' entails cognitive and other costs \citep{barca2015tracking,festinger1962theory,lepora2015embodied,ortega2013thermodynamics,pezzulo2018hierarchical,zenon2019information}}. \textcolor{colrev}{These and other cognitive biases may have some adaptive rationality in daily conditions. For example, exponential phenomena are uncommon in our daily experiences -- they are not part of our ``natural statistics''. Furthermore, circumstances that are far from us in time and space are less likely to affect us and discounting them is often safe.}

\textcolor{colrev}{Yet the unconventional circumstances of the COVID-19 pandemic challenge the adaptive rationality beyond these cognitive biases, hence exposing its limits. The most obvious example is the fact that the coronavirus can grow in an exponential manner, but people often fail to acknowledge it \citep{lammers2020correcting}. This underestimation can be catastrophic if it guides our moral decisions}: if travelers from Milan evaluated the costs of their decisions in terms of a linear growth of the virus, they would have severely underestimated them. To give a measure of such underestimation, the total cases reported in Campania on April 5th are 2960, whereas a linear projection from the data available on March 8th (the day of the escape from Milan) would result in 292 total cases -- the difference is a factor of ten.


\section{What next? Making scientific knowledge and tools more available and understandable}
%
%
%
A similar underestimation of risks may help explaining \textcolor{colrev}{other risky choices documented by the media during the early days of the spread of the virus, such as the case of individuals deciding to leave quarantine for a walk and public authorities deciding to postpone lockdowns of companies, towns (e.g., the case of Ischgl in Austria) or countries.} When the problems with COVID-19 were already apparent in China, and there were already indications that the virus was spreading in the north of Italy, the dangers were initially neglected. This is evinced by the widespread slogan ``Milano does not stop'' (in Italian, ``Milano non si ferma'') and invitation to keep going as usual by prominent politicians \citep{pisano2020lessons}.
%
The pattern of results we discussed 
defies our intuitions about the societal costs of our individual moral decisions -- and requires increased awareness and sense of individual responsibility. 

\textcolor{colrev}{During the pandemic, the media are giving great prominence to people's individual responsibility, to data and mathematical predictions about the spread of the virus, and to the opinion of scientific experts, such as virologists and modellers. However, public appeals may fail to reach their full potential (i.e., influence people's decisions) if they do not match the ways people process and evaluate information. The mere exposure of a large body of scientific knowledge may have a limited impact if it is not accompanied by measures that increase our understanding of the processes that guide our decisions (moral or otherwise) in challenging situations; and our awareness of the potential societal costs of our individual choices. We argue that increased attention to the processes that govern decision-making may be important for both those who have to develop efficacious communication strategies and for the large public interest.}

\textcolor{colrev}{The extent to which being more aware of our decision strategies and their limitations is helpful remains to be fully established. However, a study suggests that correcting the misperception of exponential coronavirus growth may support safer social interactions \citep{lammers2020correcting}. Another strategy consists in making people more aware of our decision making strategies and most importantly, what renders critical decisions difficult and psychologically challenging; for example, the fact that they often imply a collision of equally deep and non-negotiable values (e.g., in the case of lockdown decisions, safety versus freedom). Even if increased awareness of the decision processes might not significantly improve the decision itself, it could at least mitigate the associated stress and the usual inertia resulting from the selection between equally negative outcomes \citep{alison2015interoperability,shortland2020moral}. A further important strategy would be rendering the tools (statistical or otherwise) that support scientific thinking and prediction more transparent to the public. An increased understanding of the tools that experts and policy-makers use to make predictions and to design interventions may contribute to raise public engagement and shift people's attitudes towards collective decisions (e.g., decisions about lockdowns) from an uncritical delegation to public authorities and experts to a real participatory process -- which may in turn increase trust and the sense of personal responsibility. Finally, two other important dimensions that render choices complex in our information-rich world are the potential information overload and the spread of misinformation usually associated to important societal issues, including the pandemic \citep{del2016spreading,cinelli2020covid}. Helping people develop good strategies to select and assess the reliability of information sources is critical to ensure the success of social policies and interventions (e.g., vaccination campaigns). While these are clearly challenging -- if not over-rationalistic and utopian -- objectives and can only be pursued at slow (educational) timescales, they may pay large dividends in the long run.}

\textcolor{colrev}{In sum, the pandemic has created a global crisis at multiple levels (e.g., medical, economic and psychological) that our societies are trying to mitigate. The incapacity to deploy adaptive decisions (at individual and societal levels) in critical conditions is an important limiting factor when dealing with the spread of the virus \citep{berenbaum2021covid}. To mitigate this pressing problem, and to prepare more efficiently for future crises, we need novel educational strategies aiming at making the most relevant knowledge from cognitive and social sciences available to the larger public, as well as providing the education required for a better public understanding of the statistical tools that are deployed in making  more informed decisions.
Finally, along with our other institutions, also the academic system would benefit from an increased awareness of the cognitive challenges we face, in order to deal more appropriately with future crises as well as to ensure transparency, rigor and public trust \citep{ferreira2020decline}.}

\section{Material and Methods}

\begin{figure}[htbp]
\centering
\includegraphics[width=0.6\textwidth]{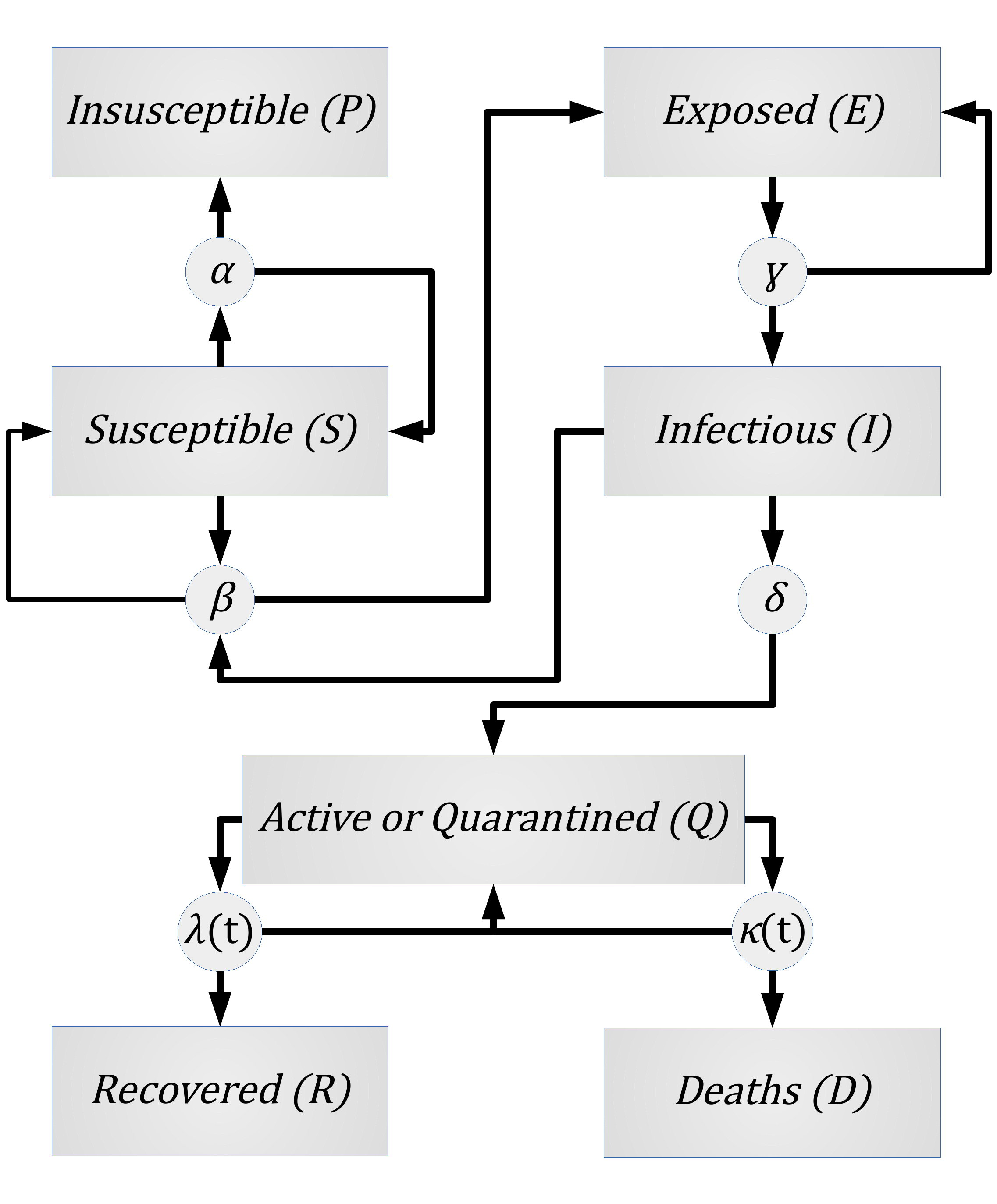}\caption{Diagram of the interactions between the modules and variables of the generalised SEIR model used for the simulations. The variables updates are computed using Eq.~\ref{eq:GSEIR}. See the main text for more details.
\label{fig:model}}
\end{figure}

The \emph{Susceptible Exposed Infectious Recovered} (SEIR) model has enjoyed considerable success in the estimation and prediction of the diffusion of COVID-19 \citep{tang2020estimation,hou2020effectiveness}. A generalised SEIR model was able to successfully characterize the outbreak of COVID-19 in Wuhan at the end of 2019 \citep{peng2020epidemic}. We adapt the same approach to the case of the ``escape from Milan'', modelling seven different variables: the number of the \emph{susceptible ($S$)} cases, \emph{insusceptible ($P$)} cases, \emph{exposed ($E$)} cases (had contact with an infected person, but not infectious), \emph{infectious ($I$)} cases (with infectious capacity but not yet quarantined), \emph{active} or \emph{quarantined} ($Q$) cases (confirmed and infectious), \emph{recovered ($R$)} cases and \emph{closed} cases ($D$) (or \emph{deaths}). The relations among the variables are shown in Fig~\ref{fig:model}.

The update of the variables is expressed by a system of first order differential equations:
\begin{equation}
\left\{ \begin{array}{ccl}
\cfrac{dS(t)}{dt} & = & -\beta\cfrac{S(t)I(t)}{N}-\alpha S(t)\\
\cfrac{dE(t)}{dt} & = & \beta\cfrac{S(t)I(t)}{N}-\gamma E(t)\\
\cfrac{dI(t)}{dt} & = & \gamma E(t)-\delta I(t)\\
\cfrac{dQ(t)}{dt} & = & \delta I(t)-\lambda(t)Q(t)\\
\cfrac{dR(t)}{dt} & = & \lambda(t)Q(t)\\
\cfrac{dD(t)}{dt} & = & \kappa(t)Q(t)\\
\cfrac{dP(t)}{dt} & = & \alpha S(t)
\end{array}\right.\label{eq:GSEIR}
\end{equation}
where 
$N=S+P +E +I +Q+R+D$ is the total population in a certain region,  
$\alpha$ is the protection rate, 
$\beta$ is the infection rate, 
$\gamma^{-1}$ is the average latent time,  
$\delta^{-1}$ is the average quarantine time, 
$\lambda(t)$ is the cure rate and 
$\kappa(t)$ is the mortality. 
\color{colrev}We used data about the COVID-19 outbreak in Campania and Lazio and Lombardia released by the 
Dipartimento della Protezione Civile (the Italian governmental  structure  that  deals  with  the  prediction,  prevention  and  management  of emergency  events),  which  can  be  downloaded  from  the  official  public  github repository  (https://github.com/pcm-dpc/COVID-19). \color{black} 
Starting from the available data, it is possible to regress the parameters of the model $\boldsymbol{\theta} = (\alpha, \beta, \gamma, \lambda(t_0), \kappa(t_0))$

Total cases $C$ are computed as the sum of three variables ($C=Q+R+D$). We simulated COVID-19 cases in a critical period of one month, starting from the weekend of the ``escape from Milan'' (March 7th-8th). First, we computed the model parameters averaged in a time-window $tw$ (8 days), from March 16th to March 23rd, for each region separately. Then, we used the resulting $\boldsymbol{\theta}_{tw}$ to simulate the next days. Predictions of total cases, active cases and deaths from March 24th ($d_0$) to April 5th ($d_{end}$) were made using three different instances $\mathcal{M}$ of the SEIR model:

\begin{itemize}
\item $\mathcal{M}$-\emph{without-travelers} with parameters $\boldsymbol{\theta}_{tw}$, regressed using data in $tw$.
\item $\mathcal{M}$-\emph{with-travelers}) with the same parameters $\boldsymbol{\theta}_{tw}$, but after a manual increase (from 1 to 100, in 100 different simulations) of the initial number of active cases ($Q$).
\item $\mathcal{M}$-\emph{ground-truth} with parameters $\boldsymbol{\theta}_{ground}$, regressed using all the available data, from March 16th to April 5th. This instance is used as a control, as it shows the best fit, when endowed with all the available data.
\end{itemize}

 The accuracy of the predictions is measured by normalized mean square error (NMSE) computed between actual total cases $T$, active cases $A$ and deaths $K$: 
 
\begin{multline}
NMSE_{\mathcal{M}}=
\frac{\sum_{td=d_0}^{d_{end}}(C_{td} - T_{td})^2}{N_T} +
\frac{\sum_{td=d_0}^{d_{end}}(A_{td} - Q_{td})^2}{N_A} + 
\frac{\sum_{td=d_0}^{d_{end}}(K_{td} - D_{td})^2}{N_K}
\label{eq:rms}
\end{multline}

The normalization factors $N_T$, $N_A$ and $N_K$ measure the difference in variance of the set of data to make the errors comparable and are computed respectively as
\begin{multline}
    \hspace{1em}
    N_T=\sum_{td=d_0}^{d_{end}}(T_{td} - \bar{T})^2
    \hspace{3em}
    N_A=\sum_{td=d_0}^{d_{end}}(A_{td} - \bar{A})^2
    \hspace{3em}
    N_K=\sum_{td=d_0}^{d_{end}}(K_{td} - \bar{K})^2\hspace{4em}
\end{multline}
where $\bar{T}$, $\bar{A}$ and $\bar{K}$ are the mean values of the variables computed on the dataset. 
Simulation results (NMSE values) are summarized in Table~\ref{tab:nmse}.
\textcolor{colrev}{Note that we selected Campania, Lazio and Lombardia for our simulations because they are the most populated regions of Italy and those for which more data are available. The source code to reproduce the figures can be downloaded from \href{https://github.com/donnarumma/MORAL\_COVID19}{https://github.com/donnarumma/MORAL$\_$COVID19}.}
\color{black}

\section*{Acknowledgements}
This research received funding from the European Union’s Horizon 2020 Framework Programme for Research and Innovation under the Specific Grant Agreement Nos. 785907 and 945539 (Human Brain Project SGA2 and SGA3) to GP, the European Research Council under the Grant Agreement No. 820213 (ThinkAhead) to GP.

\begin{table}
\begin{center}
\begin{tabular}{|c|c|c|c|}
\hline 
Instance of SEIR model & Lombardia & Lazio & Campania\tabularnewline
\hline 
\hline
without-travelers & 1.2523 & 3.0511 & 8.4054\tabularnewline
\hline 
with-travelers (best result) & ---- & 0.0145 (36) & 0.5967 (37)\tabularnewline
\hline 
ground truth & 0.1442 & 0.0078 & 0.0718 \tabularnewline
\hline
\end{tabular}
\caption{NMSE values of three instances of SEIR model (\emph{ground-truth}, \emph{without-travelers}, \emph{with-travelers}) in the three Italian regions. For \emph{with-travelers} case, only the best results are shown (with 36 and 37 active cases added to Lazio and Campania, respectively).}
\label{tab:nmse}
\end{center}
\end{table}

\bibliographystyle{cas-model2-names}

\bibliography{references}

\newpage

\newpage

\newpage

\newpage

\newpage

\end{document}